%% file: cbsoftTools.tex
\documentclass[12pt]{article}

\usepackage{sbc-template}

\usepackage{graphicx,url}
\usepackage{multirow}

\usepackage[latin1]{inputenc}  
\usepackage{color}

\title{ModularityCheck: A Tool for Assessing Modularity using Co-Change Clusters }

\author{Luciana Lourdes Silva\inst{1,2}, Daniel F\'elix\inst{1}, Marco T\'ulio Valente\inst{1}, Marcelo de A. Maia\inst{3} }

\address{Department of Computer Science -- Federal University of Minas Gerais (UFMG)
\nextinstitute
  Federal Institute of Minas Gerais -- IFMG
\nextinstitute
  Faculty of Computing -- Federal University of Uberl\^andia  
  \email{\{luciana.lourdes, dfelix, mtov\}@dcc.ufmg.br, marcmaia@facom.ufu.br}
}

\begin{document} 

\maketitle

\begin{abstract}
It is widely accepted that traditional modular structures suffer from the dominant decomposition problem. 
Therefore, to improve current modularity views, it is important to investigate the impact of design decisions concerning modularity in other dimensions, as the evolutionary view. 
In this paper, we propose the ModularityCheck tool to  assess package modularity using co-change clusters, which are sets of classes that usually changed together in the past. 
Our tool extracts information from version control platforms and issue reports, retrieves co-change clusters, generates metrics related to co-change clusters, and provides visualizations for assessing modularity. We also provide a case study to evaluate the tool.

http://youtu.be/7eBYa2dfIS8

\end{abstract}

\section{Introduction}

\input{sections/introduction.tex}

\section{ModularityCheck in a Nutshell} \label{sec:firstpage}

\input{sections/overview.tex}

\section{Use Case Scenario: Geronimo Web Application Server}

\input{sections/caseStudyTool.tex}

\section{Related Tools}
\input{sections/relatedWorkTool.tex}

\section{Conclusion}
\input{sections/conclusion.tex}

\section*{Acknowledgement} This work was supported by CNPq, CAPES, and FAPEMIG.

\bibliographystyle{sbc}
\bibliography{sbc-template}

\end{document}

%% file: sections/introduction.tex
There is a growing interest in tools to enhance software quality~\cite{mylyn, zeller2005}. 
Specifically, several tools have been developed for supporting software modularity improvement~\cite{Rebelo:2014, Vacchi:2014, Bryton2008, Schwanke1991}. Most of such tools help architects to understand the current package decomposition. Basically, they extract information from the source code by using structural dependencies and the source code text~\cite{concerngraphs_tosem, concerngraphs}. 

Modularity is a key concept when designing complex software systems~\cite{power-modularity}. 
The central idea is that modules should hide important design decisions or decisions that are likely to change~\cite{parnas72}. 
Typically, the standard approach to assess modularity is based on coupling and cohesion, calculated using the structural dependencies established between the modules of a system (coupling) and between the internal elements from each module (cohesion).
Usually, high cohesive and low-coupled modules are desirable because they ease software comprehension, maintenance, and reuse. 
However, typical cohesion and coupling metrics measure a single dimension of the software implementation (the static-structural dimension). 
On the other hand, it is widely accepted that traditional modular structures and metrics suffer from the dominant decomposition problem and  tend to hinder different facets that developers may be interested in~\cite{mylyn,concerngraphs,concerngraphs_tosem}. For example, there are various effects of coupling that are not captured by structural coupling. 
Therefore, to improve current modularity views, it is important to investigate the impact of design decisions concerning modularity in other dimensions of a software system, as the evolutionary dimensions. 

To address this question, we present in this paper the ModularityCheck tool to support package modularity assessment and understanding using co-change clusters. The proposed tool has the following features:

\begin{itemize}
	\item The tool extracts commits automatically from the version history of the target system and discards noisy commits by checking with their issue reports.
	\item The tool retrieves set of classes that usually changed together in the past, which we termed co-change clusters. 
	\item The tool relies on distribution maps~\cite{ducasse2006} to reason about the projection of the extracted co-change clusters in the tradition decomposition of a system in packages. It also calculates a set of metrics defined for distribution maps to support the characterization of the extracted co-change clusters.

\end{itemize}

%% file: sections/overview.tex
\begin{figure*}[!t]
\begin{center}
\includegraphics[scale=0.3]{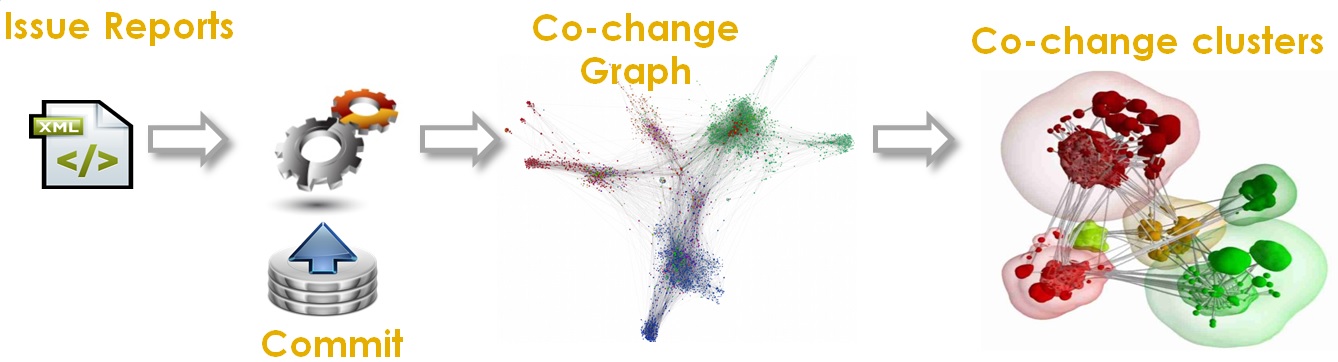}
\caption{ModularityCheck's overview.}
\label{fig:figure1}
\end{center}
\end{figure*}

ModularityCheck supports the following stages to asses the quality of a system package modularity: pre-processing, post-processing, co-change clusters retrieval, and cluster visualization. Figure~\ref{fig:figure1} shows the process to retrieve co-change clusters. A detailed presentation of this process is available in a full technical paper~\cite{modularity2014}.

In the first stage, the tool applies several preprocessing tasks which are responsible for selecting commits from version history to create the co-change graph. 
In such graphs, the vertices are classes and the edges link classes changed together in the same commits. In the second stage, a post-processing task prune edges with small weights from the co-change graphs. 
After that, the co-change graph is automatically processed to produce a new modular facet: co-change clusters, which abstract out common changes made to a system, as stored in version control platforms. Therefore, co-change clusters represent sets of classes that changed together in the past. 
Finally, the tool uses distribution maps~\cite{ducasse2006}---a well-known visualization technique---to reason about the projection of the extracted clusters in the traditional decomposition of a system in packages. 
ModularityCheck also provides a set of metrics defined for distribution maps to reason about the extracted co-change clusters. 
Particularly, it is possible to reason about recurrent distribution patterns of co-change clusters listed by the tool, including patterns denoting well-modularized and crosscutting clusters. 

\subsection{Architecture}

ModularityCheck supports package modularity assessment of software systems implemented in the Java language. The tool relies on the following inputs: 
(i) the issue reports saved in {\tt XML} files; (ii) URL of the version control platform (SVN or GIT). (iii) maximum number of packages to remove highly scattered commits. 
(iv) minimum number of classes in a co-change cluster. We discard small clusters because they may eventually generate a decomposition of the system with hundreds of clusters. Figure~\ref{fig:figure2} shows the tool's architecture which includes the following modules:

\begin{figure*}[!t]
\begin{center}
\includegraphics[scale=0.65]{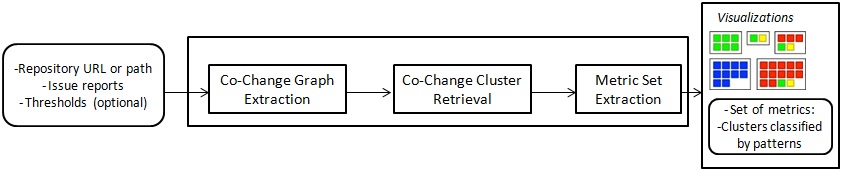}
\caption{ModularityCheck's architecture.}
\label{fig:figure2}
\end{center}
\end{figure*}

\paragraph{Co-Change Graph Extraction:} As illustrated in Figure~\ref{fig:figure2}, the tool receives the 
URL associated to the version control platform of the target system and the issue reports. When extracting co-change graphs, it is 
fundamental to preprocess the considered commits to filter out commits that may pollute the graph with noise. Firstly, the tool removes
commits not associated to maintenance issues because commits can denote partial implementations of programming 
tasks. Secondly, the tool removes commits not changing classes because the co-changes considered by ModularityCheck are defined for classes. Thirdly, commits associated to multiple maintenance issues are removed. Such commits could generate edges connecting classes modified to implement semantically unrelated maintenance tasks, which were included in the same commit just by convenience, for example. Finally, the last pruning task removes highly scattered commits, according the \textit{Maximum Scattering} threshold, an input parameter. 
Such commits usually are associated to refactoring activities, dead code removal, or changes to comment styles. The default value considered by the tool is ten packages.

\paragraph{Co-Change Cluster Retrieval:} After extracting the co-change graph, a post-processing tasks is applied to prune edges with small weights. In this phase, edges with weights less than two co-changes are removed.  
Then, in a further step, a data mining algorithm named Chameleon~\cite{Karypis1999} is performed to retrieve subgraphs with high density. The number of clusters is defined by executing Chameleon multiple times. After each execution, small clusters are discarded by the \textit{Minimum Cluster Size} threshold informed by the user. The default value considered by the tool is 4 classes, i.e., after the clustering execution, clusters with less than four classes are removed.

\paragraph{Metric Set Extraction:} The tool calculates the number of vertices, edges, and co-change graph's density before and after the post-processing filter. After retrieving the co-change clusters, the tool presents the final number of clusters and several standard descriptive statistics measurements. These metrics describes the size and density of the extracted co-change clusters, and cluster average edges' weight. Moreover, the tool presents metrics defined for distribution maps, like focus and spread. 
ModularityCheck also allows to investigate the distribution of the co-change clusters over the package structure by using distribution maps~\cite{ducasse2006}. 
In our distribution maps, entities (classes) are represented as small squares and package structure groups such squares into large rectangles. In the package structure, we only consider classes that are members of co-change clusters, in order to improve the maps visualization. Finally, all classes in a co-change cluster have the same color. 

The {\em focus} of a given cluster $q$ in relation to package structure P is defined, as follows:
\[
\mathit{focus}(q,P) = \sum_{p_i \in P}{\mathit{touch}(q,p_i)*\mathit{touch}(p_i,q)}
\]
where
\[
\mathit{touch}(p,q)= \frac{| p \cap q | }{|q|}
\]

\noindent In this definition, $\mathit{touch}(q,p_i)$ is the number of classes of cluster $q$ located in the package $p_i$ divided by the number of classes in $p_i$ that are included in at least a co-change cluster. Similarly, $\mathit{touch}(p_i,q)$ is the number of classes in $p_i$ included in the cluster $q$ divided by the number of classes in $q$.  Focus ranges between 0 and 1, where the value one means that the cluster $q$ dominates the packages that it touches. There is also a second metric, called {\em spread}, that measures the number of packages touched by $q$.

After measuring focus and spread, the tool classifies recurrent distribution patterns of co-change clusters, as follows: well-encapsulated, partially encapsulated, well-confined in packages, or crosscutting clusters. 

%% file: sections/caseStudyTool.tex
\begin{figure*}[!t]
\begin{center}
\includegraphics[scale=0.55]{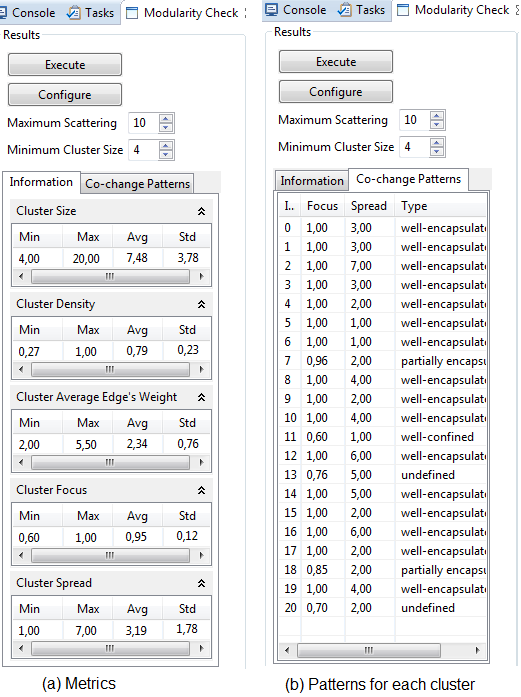}
\caption{Filters and metric results.}
\label{fig:figurea}
\end{center}
\end{figure*}

In order to present ModularityCheck, we provide a scenario of usage involving information from the Geronimo Web Application Server system, extracted during 9.75 years (08/20/2003 - 06/04/2013). Figure~\ref{fig:figurea} shows the results concerning co-change clustering.
A detailed discussion of such results is presented in technical paper~\cite{modularity2014}.

\subsection{Co-Change Extraction} First, our tool extracted 9,829 commits. We maintained the value for \textit{Maximum Scattering} as 10, i.e., the tool discarded commits changing classes located in more than ten packages. After the pre-processing tasks, only 1,406 commits were considered as useful.

\subsection{Co-Change Clustering} In the next step, small clusters are discarded by following \textit{Minimum Cluster Size} filter. The tool removed clusters with less than 4 classes, resulting in 21 clusters.
The ratio between the final number of clusters and the number of packages in the system is 0.05\%. This fact is an indication that the maintenance activity in the system is concentrated in few classes. 

Figure~\ref{fig:figurea}a shows standard descriptive statistics measurements regarding the size, density, average edge's weight of the extracted co-change clusters. 
ModularityCheck presents the size of the extracted co-change clusters, in terms of number of classes. 
The extracted clusters have $7.48 \pm 3.78$ classes in Geronimo. Moreover, the biggest cluster has a considerable number of classes: 20 classes. 
The tool also presents the density of the extracted co-change clusters. The clusters have a density of $0.79 \pm 0.23$. 
We can also analyze the average weight of the edges in the extracted co-change clusters. For a given co-change cluster, we define this average as the sum of the weights of all edges divided by the number of edges in the cluster. We can observe that the average edges' weight is not high, being slightly greater than two in Geronimo. 

\subsection{Modularity Analysis}

ModularityCheck also provides a visualization, which relies on co-change clusters to assess the quality of a system's package decomposition. Basically, this visualization allows to reveal the distribution of the co-change clusters over the package structure by using distribution maps. 
The tool also shows the standard descriptive statistics measurements regarding respectively the focus and spread of the co-change clusters. As presented in Figure~\ref{fig:figurea}a, the co-change clusters in Geronimo have high focus with the average 0.95. Regarding spread, on average the spread is 3.19. Figure~\ref{fig:figurea}b shows the focus, spread, and type of patterns for each cluster.

\subsubsection{Geronimo Results}

\begin{figure*}[!t]
\begin{center}
\includegraphics[scale=1.3]{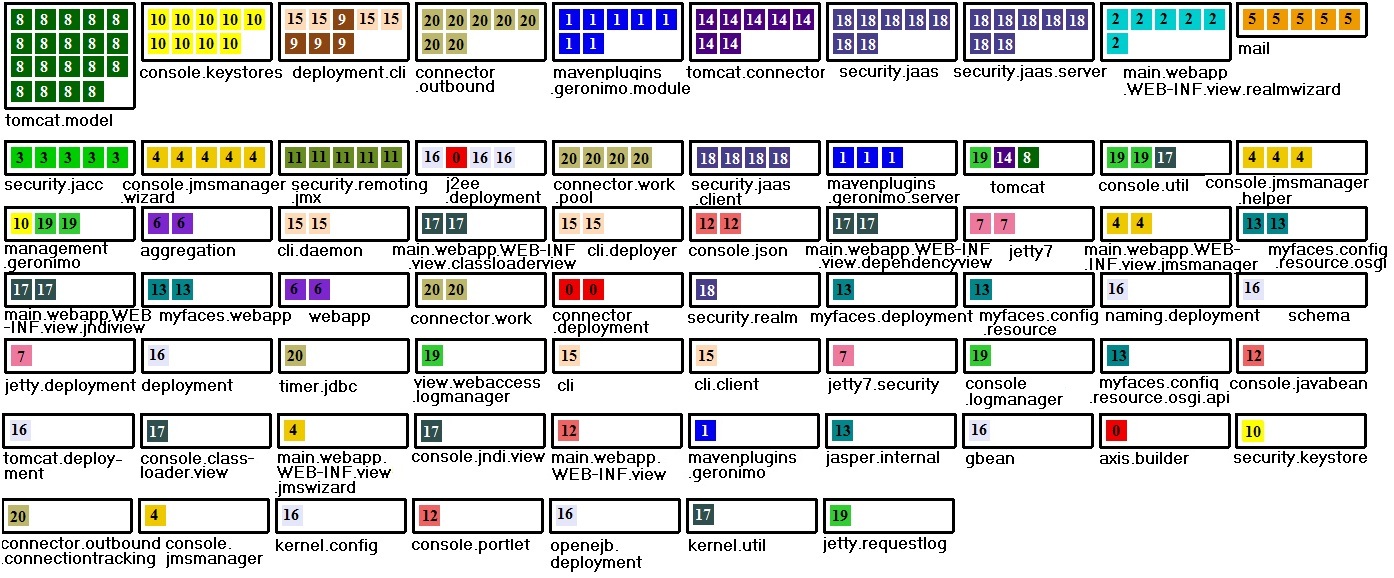}
\caption{Distribution maps for Geronimo~\cite{modularity2014}.}
\label{fig:figure3}
\end{center}
\end{figure*}

Figure~\ref{fig:figure3} shows the distribution map for Geronimo. To improve the visualization, besides background colors, we use a number in each class (small squares) to indicate their respective clusters. If we stop the mouse over a class, a tooltip is displayed with its respective name. 
The large boxes are the packages  and the text below is the package name.

Considering the clusters that are {\em well-encapsulated} (high focus) in Geronimo, we found two relevant distribution patterns:
\begin{itemize}
\item {\em Clusters well-encapsulated (focus = 1.0) in a single package  (spread = 1)}. Four clusters have this behavior. As an example, we have Cluster 2, which dominates the co-change classes in the package {\tt main.webapp.WEB\-INF.view.realmwizard} (line 1 in the map, column 9). Cluster 5 (package {\tt mail}, line 1 in the map, column 10) and Cluster 11 (package {\tt security.re\-moting.jmx}, line 1, column 3).

\item {\em Clusters partially encapsulated (focus $\approx$ 1.0), but touching classes in other packages (spread $>$ 1)}. As an example, we have Cluster 8 ({\em focus} = 0.97, {\em spread} = 2), which dominates the co-change classes in the package {\tt tomcat.model} (line 1 and column 1 in the map), but also touches the class {\tt TomcatServerGBean} from package {\tt tomcat} (line 2, column 8). 

\end{itemize}

\subsection{Practical Usage}

ModularityCheck can support software architects to assess modularity under an evolutionary view. It helps to detect co-change behavior patterns, as follows: 

\begin{itemize}

\item When the package structure is adherent to the cluster structure, localized co-changes are likely to occur, as in Geronimo's clusters. 

\item When there is no a clear adherence to the cluster structure. Our tool detects two cluster patterns that may suggest modularity flaws. The first pattern denotes clusters with crosscutting behavior, not detected in Geronimo but we could detect them in other systems presented in~\cite{modularity2014}. The second indicates clusters partially encapsulated that suggest a possible ripple effect during maintenance activities.  
\end{itemize}

%% file: sections/relatedWorkTool.tex
Zimmermann et al.~proposed ROSE, a tool that uses association rule mining on version histories to recommend further changes~\cite{zeller2005}. Their tool differs from ours because they rely on association rules and we use co-change clusters that are semantically related to a maintenance task. Furthermore, our goal is not to recommend future changes but to assess modularity using distribution maps to compare and contrast co-change clusters with the system's packages.

ConcernMapper~\cite{Robillard2005} is an Eclipse Plug-in to organize and view concerns using a hierarchical structure similar to the package structure. However, the concern model is created manually by developers and the relations between concerns are typically syntactical and structural. On the other hand, in our tool, the elements and their relationships are obtained by mining the version history. Particularly, relationships express co-changes and concerns are retrieved automatically by clustering co-change graphs. 

Wong et al. presented CLIO, a tool that detects and locates modularity violations~\cite{Wong:2011}. CLIO compares how components should co-change according to the modular structure and how components usually co-change retrieving information from version history. A { \em modularity violation} is detected when two components usually change together but they belong to different modules, which are supposed to evolve independently. 
CLIO identifies modularity violations by comparing the results of structural coupling with the results of change coupling. 
They compare association rules and structural information to detect modularity violations. On the other hand, we retrieve co-change clusters and use distribution maps to reason about the projection of the extracted clusters in the traditional decomposition of a software system in packages. 

Palomba et al.~proposed HIST, a tool that uses association rule mining on version histories to detect the following code smells: Divergent Change, Shotgun Surgery, Parallel Inheritance, Blob, and Feature Envy~\cite{PalombaASE:2013}. HIST bases on changes at method level granularity. For each smell, they defined a heuristics that relies on association rules discovery or that analyzes co-changed classes/methods for detecting bad smells. In contrast, our goal is not to detect code smells but to assess package decomposition using co-change clusters.

%% file: sections/conclusion.tex
In this paper, we proposed a tool to assess modularity using evolutionary information. The tool extracts commits automatically from version histories and filter out noisy information by parsing issue reports. After that, the tool retrieves co-change clusters, a set of metrics concerning clusters, and provides a visualization based on distribution maps. The central goal of ModularityCheck is to detect classes of the target system that usually change together to help on assessment of the package modular decomposition. 
Moreover, the co-change clusters can also be used as an alternative view during maintenance tasks to improve the developer's comprehension of the their tasks.
The ModularityCheck tool is publicly available at: \textbf{aserg.labsoft.dcc.ufmg.br/modularitycheck}